\documentclass[fleqn,usenatbib]{mnras}

\usepackage{newtxtext,newtxmath}

\usepackage[T1]{fontenc}
\usepackage{ae,aecompl}

\usepackage{graphicx}	
\usepackage{amsmath}	
\usepackage{makecell}
\usepackage{multirow}
\usepackage{subcaption} 
\usepackage{graphbox}
\usepackage{float}

\usepackage{xcolor}

\title[Dynamical Dark Energy Properties Hidden in the Dark  Halos and Matter Voids]{Dynamical Dark Energy Properties Hidden in the Dark Matter  Halos and Voids}

\author[Aghileh ebrahimi et al.]{
Aghileh S Ebrahimi,$^{1,2}$\thanks{E-mail: aghileh.ebrahimi@gmail.com}
A. Vafaei Sadr$^{3,4}$
S. Tavasoli$^{2,5}$
\\
$^{1}$Department of Physics, University of Kashan, Ravand, Kashan, Iran.\\
$^{2}$School of Astronomy, Institute for Research in Fundamental Sciences (IPM), Tehran, Iran.\\
$^{3}$Department of Physics, Institute for Research in Fundamental Sciences (IPM), Tehran, Iran.\\
$^{4}$African Institute for Mathematical Sciences, 6 Melrose Road, Muizenberg, South Africa\\
$^{5}$Physics Department, Faculty of Science, Kharazmi University, Tehran, Iran}

\begin{document}

\label{firstpage}
\pagerange{\pageref{firstpage}--\pageref{lastpage}}
\maketitle

\begin{abstract}
In this paper, we analysed  the halos and voids properties of a GR-based N-body simulation carried out at redshifts $z=0.0$  and $ z=0.8$ as differences between dynamical dark energy models (namely PL and CPL) with respect to $\Lambda CDM$. Analysing the halos demonstrates that both models, PL and CPL, behave like $\Lambda CDM$, despite the velocity dispersion of halos was more sensitive to the dynamical dark energy model. In addition, a void finder was developed to extract the properties of voids from simulated data. Further statistical model on voids confirms that the PL model produces larger voids. In summary, our novel simulation demonstrates void properties are better than halo properties in discriminating between dark energy models. Hence, the results suggest to make more use of the properties of voids in future studies of discriminating dynamical dark energy models. 
 
\end{abstract}

\begin{keywords}
Dynamical dark energy, Halos properties, Voids properties, N-body simulation, Structure formation.
\end{keywords}



\section{Introduction}
The cosmological constant was historically introduced to allow for obtaining a static solution of the Fridman equations, but it received more attention after the discovery of the acceleration of the expansion\citep{Riess:1998cb,Perlmutter:1998np}.
Various scientific projects  such as the DESI \cite{flaugher2014dark}, the WiggleZ \cite{parkinson2012wigglez},the Destiny \cite{benford2006destiny} and the Planck \cite{Ade:2015fva} were carried out not only to assess the source of expansion but also to achieve desired observational accuracy \citep{Cheng:2014kja}.  However, the $\Lambda CDM$ model is a simple and well defined model which is consistent
with the majority of current cosmological observations. However, it suffers from some problems such as the sharp transition from the cold dark matter era to $\Lambda$ the dominant epoch and the fine-tuning \citep{Martin:2012bt,Kunz:2012aw}. Besides, there are some reports on flowing tension: the excessive congestion of the matter and the missing satellites puzzle \cite{Moore:1999nt,Klypin:1999uc}, the cusp-core problem \citep{Navarro:1995iw} and in reliable observational projects run by the $Planck$ collaboration, the observed clusters are fewer than what was expected \citep{Ade:2015fva}. There is a tension between the value of the Hubble constant from local distance indicators as the late time data and the angular scale of fluctuations in the Cosmic Microwave Background in early time data \cite{freedman2017cosmology,mortsell2018does,ebrahimi2020tension}. Due to the mentioned problems, $\Lambda$ can be considered as a variable quantity rather than a constant value.
Hence, reconstruction of the equation of state  based on observations \citep{Huterer:1998qv} and model parameters estimation \citep{Nair:2013sna} are performed to alleviate the problems.\\
The important effects of dark energy on structures are in the red-shift evolution of the linear growth factor and the evolution of perturbations which are sensitive to the dark energy equation of state. The statistical analysis of dark energy properties mainly focused on highly dense regions such as the galaxy number density, two-point correlation function, and gas galaxy clusters \citep{McClintock:2018bxh,benson2018wfirst}. The cosmological peculiar velocity of dark matter halo carries significant information on the theory of gravity and dark energy in large scales \cite{Nusser:2012jx}. Furthermore,  the evolution of dark energy filed value produces a term which decreases or increases the peculiar velocity \citep{Farrar:2003uw}. The suppression or enhancement peculiar velocity depends on whether $ w>-1$ or $w<-1$ compared with $\Lambda CDM$ \cite{Xu:2013jma}.  Moreover, extra sensitivity of voids to the dark energy equation of state has been determined \citep{2010MNRAS.403.1392L,lavaux2012precision}. Recent research points out that cosmic voids not only show a key component of the mass distribution but also are one of the cleanest demonstrations of the dark energy's nature \citep{Bos:2012wq}. Voids provide an opportunity to investigate dark energy properties without any new survey.  In addition, it has been demonstrated voids abundance noticeably do tighter constraints on free parameters of the dark energy equation of state \cite{pisani2015counting}. Using voids as a cosmological probe can lead to some complications as the void definition and the void finding since voids are not well-defined and unbounded systems. The void has been classified into three categories: firstly, it determines the empty regions of the universe, secondly,  one tries to find voids as a geometrical structure in dark matter, and thirdly, it checks unstable points in the distribution of galaxies \citep{2010MNRAS.403.1392L,Colberg:2008qg}.\\
Releasing of the large public catalogue \citep{2012MNRAS.421..926P,Sutter:2012wh,Sutter:2012tf} from the Solan Digital Survey (SDSS)  provides a comparison of the simulation with data and highlights the importance of N-body simulations for studying void properties.\\
 Moreover, several attempts has been done to extract voids on cosmological data. Two methods have been developed to extract voids in Dark Energy Survey (DES). Further, the relation of mass and galaxies profile of cosmic voids base on N-body simulation and 3D distribution of galaxies has been studied \cite{fang2019dark}. In addition, an algorithm of void finder developed and applied to photometric DES-SV data in the redshift range $0.2 < z < 0.8$ \cite{sanchez2016cosmic}. A void size function model has been developed on simulation of dark matter halo which calibrated on the halo catalogue and the method is a critical step in applying to real data \cite{contarini2019cosmological}.\\
The cosmic voids give us an opportunity to study massive neutrino. Neutrino free steaming in the voids is more easily rather than the dark matters and the baryons. It has been shown that the characteristics of the voids are affected by neutrino through DEMNUni simulations \citep{schuster2019bias}. Besides, the theoretical description of the void size function instead of direct simulation of halo catalogue in DEMNUni simulations have been tested \cite{verza2019void}.\\
Several existing void finder algorithms have been tested to check the potential power of discrimination between General Relativity and modified gravity. They found the voids profile of f(R) theory is more underdense rather than General Relativity. They further investigated the potential of each void finder to test f(R) in the near future survey as Euclid and LSST \cite{cautun2018santiago}.

There are a vast number of void finder algorithms based on finding the minimum underdensity working on unsmooth particles \cite{neyrinck2008zobov,sutter2015vide} or a well-known technique
for the segmentation of images \cite{platen2007cosmic}. \citep{Colberg:2008qg} presents the systematic comparison of 13 voids finders using particles and halos and found that various methods are in agreement. \\

At present, N-body simulations for studying the structure formation still employ the Newtonian approximation \citep{springel2001gadget,Schmidt:2009sg,Li:2011vk} which impose restrictions on the nature of dark energy and dark matter. Newtonian N-body simulations are non-relativistic \cite{springel2005cosmological} and could not compute the essential relativistic nature of scalar fields \cite{Noller:2013wca}. To predict and compare the upcoming observations from galactic and cluster scale in non-linear regimes, N-body simulations are required to solve scalar field equations \cite{llinares2014isis}. Hence, quasi statistic approximation was added to Newtonian simulations \cite{llinares2014isis,puchwein2013modified}, despite it leads to high computational time cost \cite{brax2013systematic,Noller:2013wca}. Furthermore, Newtonian simulations work on the background evolution of Universe \cite{pollina2016cosmic} and they fail to describe the dark energy clustering in perturbation level which is an important aim of upcoming surveys such as Euclid \cite{Laureijs2011gra}. Gevolution is the first N-body code that considers all the six degrees of freedom in the metric and solves the geodesic equation. It takes into account the relativistic potential which is sensitive to the dark energy equation of state and it is more appropriate for the dark energy scenario \citep{Adamek:2015eda}. Recently, a relativistic N-body code for the clustering of dark energy has been developed \cite{hassani2019k}. It is based on the Gevolution and focuses on k-essence dark energy.

In this paper, we focus on N-body simulation of the dynamical dark energy models based on General Relativity with Gevolution and use the void finder which works with an unsmooth particle distribution without any prior spherical shape.We also utilize the halo finder analysis to extract the statistical properties of the dark matter halo for the models. Then, we employ the void finder based on the Alikio and Mahonen method to explore voids' size and density distributions for the dynamical dark energy and $ \Lambda CDM $ models. 
The chosen models in this study are Chavelier-Polarski-Linder (CPL) and Power Law (PL). Measurement of free parameters of the model from combination of SNIa+BAO+HST+\textit{Planck}TT+LSS observation data sets are obtained \cite{Ebrahimi:2018eod}. The model has quite a different equation of state from $\Lambda$CDM. The CPL equation of state remains always below of $w=-1$ and the PL equation of state crosses $w=-1$. 
Due to the crossing behavior of PL, the model has a positive equation of state at early times and enhances the matter component of the early Universe, leading it to naturally behaves similarly to coupled dark energy- dark matter models without introducing any extra coupling parameter.In section \ref{theory}, we present the dark energy models. We intend to simulate the reported observational constraints on their free parameters. In section \ref{simu}, we investigate the properties of N-body simulation. Then, the results of the halo finder and void finder  are presented in section \ref{result}. The  discussion are given in section \ref{discussion}. Finally, we conclude the main findings in section \ref{conclusion}.

\section{THEORY}\label{theory}
Dark energy properties are often characterized  by  the equation of state ($ w(z) $),  which introducing different parameterization of equation of state. It allows us to illustrate the Universe evolution. Also, the sound  speed ($ c_{s}^{2} $) causes clustering of dark energy, and anisotropic  pressure ($ \sigma $). Anisotropic  pressure  appear when the first order of the perturbation is considered.  Dynamical dark energy models can be distinguished from $\Lambda$CDM through  a variable equation of state and non-zero sound speed. The anisotropic pressure  is zero in the dark energy models we simulated. 
The models that are considered in this study are as following:

\subsection{CPL model}
Reconstruction of equation of state is a viewpoint which extract the property of dark energy as a component of the Universe. Equation of state can parameterized by using free parameters which its number depends on theoretical or phenomenological approach. Fist equation of state  was proposed by \cite{linder2003exploring} base on linear series of redshift  and suffer divergence at high redshift. Then, Chavelier-Polarski-Linder model (CPL) has been proposed  to explore the dynamical behavior of  dark energy with the following equation of state \citep{Chevallier:2000qy,Linder:2002et}:
\begin{equation}
\label{eq:wcpl}
w_{\rm CPL}(z)=w_{0} + w_{1}\dfrac{z}{1+z}.
\end{equation}
where $w_0$ and $w_1$ are the free parameters of the model. It covers many scalar field  EoS with high accuracy. It has manageable two dimensional phase space, bounded at high redshift and widely used \cite{vazquez2012reconstruction,sahni2008two,bengaly2020null,bengaly2020hubble}. However, it  suffers from a divergence at $z=-1$. The CPL model is in subclass of quintessence and its Lagrangian ca be found \cite{ma2011probing}.\\

\subsection{PL model}
Power law (PL) model is parametrized to solve the fine tuning problem. The PL model does not suffer the fine tuning problem because the ratio of the dark energy density to the matter density is not sensitive to the free parameters  of the model and asymptotically goes to zero at the early epochs.  Also, the model solves the age of old star problem known as cosmic age crisis. The PL model equation of state has a crossing $w=-1$ behaviour which leads the dark matter effect at early universe and behaves same as coupled dark matter-dark energy models without any coupling parameter. The equation of state of the model is given by  \citep{Rahvar:2006tm}:
\begin{equation}\label{eq:wpl}
w_{\rm PL}(z) = \frac{w_{0}}{(1+z)^{\alpha}}[1 -\alpha \ln\left(1+z \right)].
\end{equation}
where $w_{0} $ and $\alpha $ are  the model's free parameters. The Lagrangian of the model can be found in \citep{Rahvar:2006tm}.\\
The observational  constraints on  the free  parameters of  the models  that  are the subject of our interest in this work are given in  Table \ref{obsevation}.

\begin{table}
	\begin{center}
		\centering
		\caption{\label{obsevation}  Observational constraints  on the free  parameters of the PL and CPL models using SNIa+BAO+HST+\textit{Planck}TT+LSS data employing the Markov Chain Monte Carlo method. }
		\begin{tabular}{|l|c|c|}
			\hline
			\hline Parameter& PL & CPL \\
			\hline $w_0$&$-1.3799^{+0.0036}_{-0.0028}$& $-1.08045^{+0.00041}_{-0.00062}$\\
			\hline $w_1$&- & $-0.12190^{+0.00050}_{-0.00030}$\\
			\hline $\alpha$&$ 0.1013\pm 0.0031$ & -\\
			\hline $\Omega_{DM}$&$0.1195\pm 0.0013$ &$0.1193^{+0.0014}_{-0.0017}$\\
			\hline $\Omega_{DE}$&$0.6862\pm 0.0078$ &$0.688^{+0.010}_{-0.0079}$\\
			\hline $H_0$&$67.36\pm 0.56$ & $ 67.48^{+0.71}_{-0.57}$\\
			\hline
			\hline
		\end{tabular}
	\end{center}
\end{table}

\begin{figure}
	\begin{center}
		\includegraphics[width=0.32\textwidth]{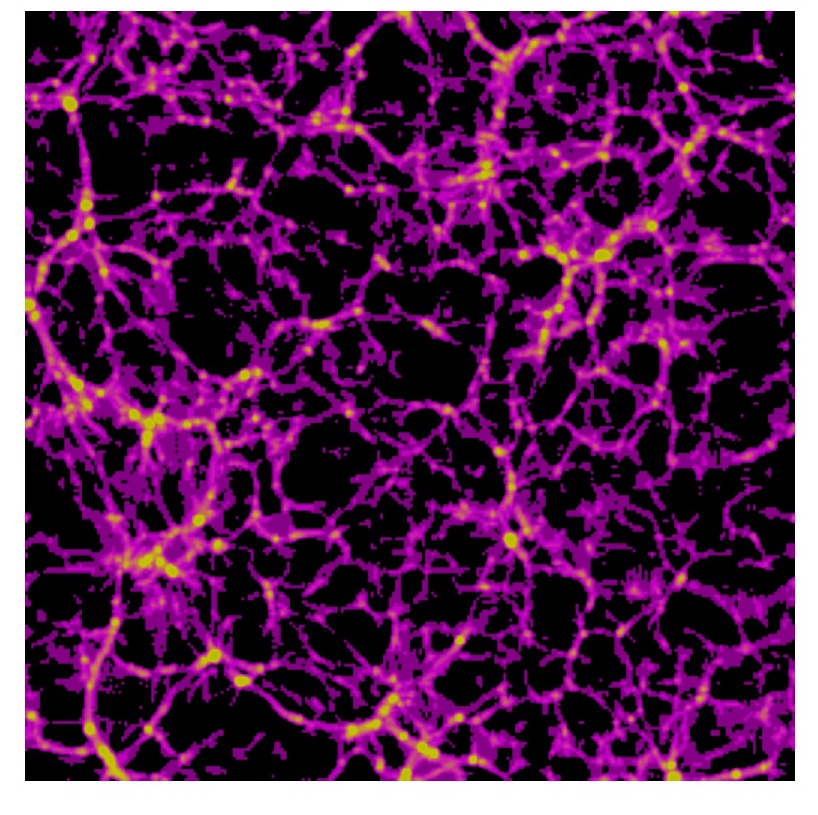}
		\includegraphics[width=0.36\textwidth]{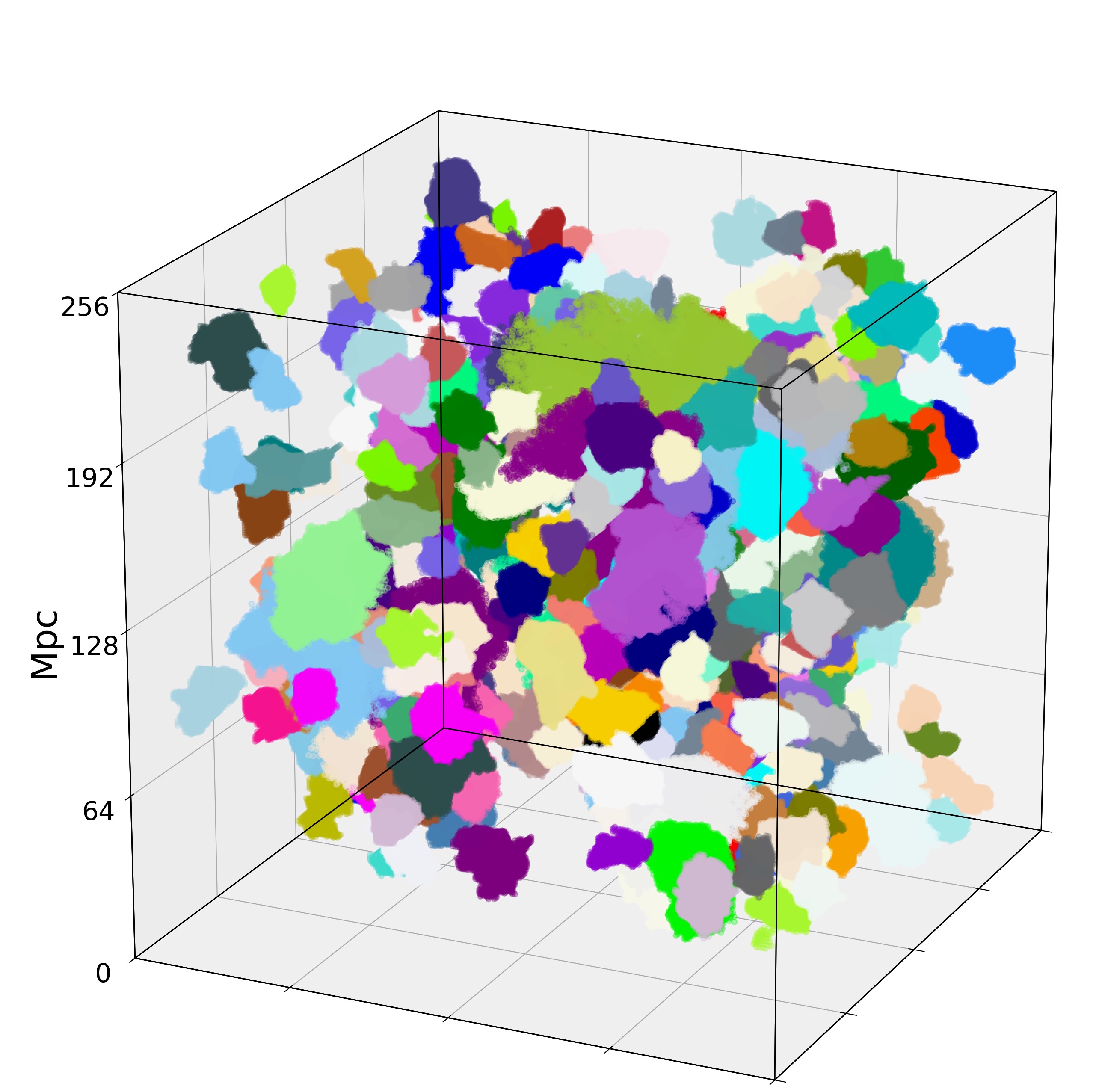}
	\end{center}
	\caption{  The top panel shows the map density of dark matter through  N-body simulation of  $\Lambda CDM$ by Gevolution at $z=0$. the side size is  256 Mpc/h. The bottom panel is the voids visualization of  $\Lambda CDM$ based on the simulation at $z=0.0$ }
	\label{fig:simulation}
\end{figure}
\begin{figure*}{h}
\centering
   \includegraphics[width=0.9\linewidth]{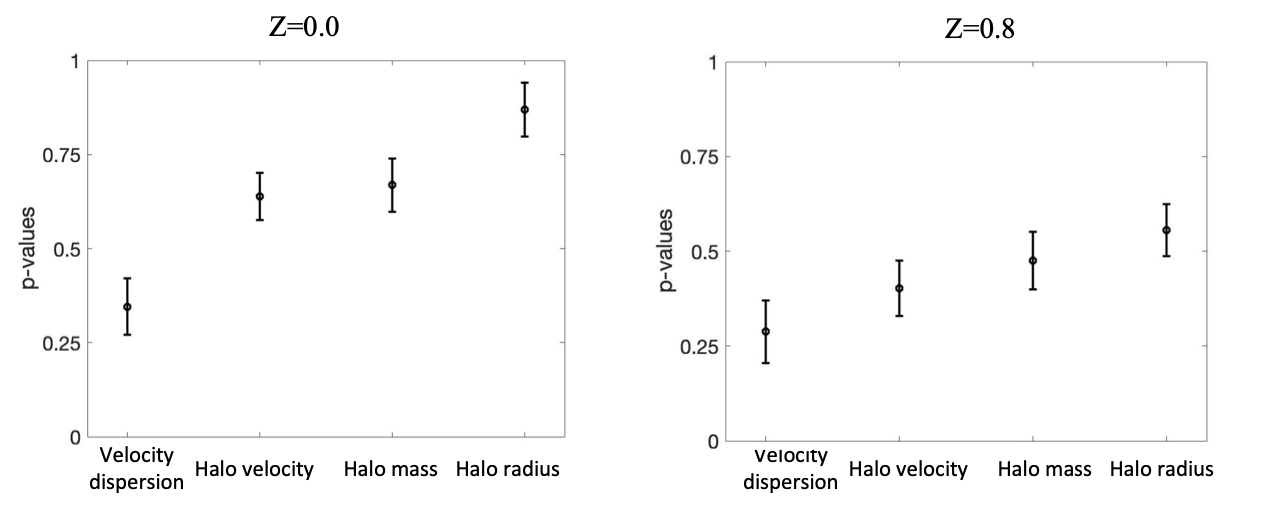}\hfil
\caption{ Plot shows the p-values of the Kruskal-Wallis test that compared the median of three models for velocity dispersion, halos velocity, halos mass and halo radius through 10 simulation boxes in two redshifts $z=0.00$ (left panel) and $z=0.8$ (right panel), both at 1 $\sigma$ confidence level.
}
\label{copmarsim}
\end{figure*}
 \section{SIMULATIONS}\label{simu}
Upcoming surveys will observe the Universe with higher accuracy and more precision in the non-linear regime \cite{Adam:2015rua,Carilli:2004nx}, namely, the Euclid will accomplish a spectroscopic redshift survey of 50 million galaxies over a volume 500 times larger than the SDSS. The spectroscopic survey will be received with a redshift accuracy $ \sigma_{z}/(1+z) \leqslant 0.001$.  Also, observing galaxies are over $75\%$  of the lifetime of the Universe. Furthermore, the precision of  photometric redshifts for these galaxies reaches $ \sigma_{z}/(1+z) < 0.05$ \cite{Laureijs2011gra}. Despite Non-linear scales are challenging, they are worth working on through N-body simulations. 
The various simulation strategies address different problems in this regime depending on the goal of the study.
In general, N-body simulations are based on Newtonian gravity or post-Newtonian gravity \citep{springel2001gadget,Schmidt:2009sg,Li:2011vk}.  They work on the background evolution of the Universe and fail to describe dark energy as a scalar field and its perturbation.
In 2016, \citep{Adamek:2016zes} introduced {\it Gevolution} which is a N-body simulation code based on General Relativity framework. The code is based on a weak field expansion of General Relativity and
calculates all six metric degrees of freedom in Poisson gauge. It addresses a more general solution among the rest of the structure formation simulations. {\it Gevolution} is an appropriate choice for studying models with dynamical dark energy or with modified gravity and it is claimed to be computationally fast enough. There are several comparisons about the speed in \cite{Adamek:2016zes}.

We modified {\it Gevolution} at the background evolution according t0 the mentioned dynamical dark energy models. The side of simulation boxes is 256 Mpc/h and contains $512^{3}$ particles using 1 Mpc/h resolution for fields taking 265 CPU hours of execution time. To assess the homogeneity of the simulation, the simulation was run 10 times with different number of seeds. Then, we applied halo finder algorithms (is explained below) to particles of 10 boxes and analysed the p-value of the dark matter halos properties. Finally, we performed the void finder analyse and compared the properties of halos and voids of the dark energy models with respect to $\Lambda CDM$.

\section{RESULTS}\label{result}

\subsection[Rockstar Halo Finder Analysis]{Rockstar Halo finder Analysis}
Since the dark matter halos carry a significant information regarding cosmology, to compare dark matter statistics with void results we analysed dark matter halo properties distribution employing the Rockstar (Robust over density Calculation using K-Space topologically adaptive refinement) halo finder for the models \cite{behroozi2012rockstar}. The Rockstar is an algorithm to identify the dark matter halos and their substructures. It is based on the adaptive hierarchical refinement of friends-of-friends groups in seven dimensions  (phase-space+time). This algorithm is very good at the substructure recovery. The Rockstar separates particles to friends-of-friends
groups at the very early step then performs subgroup detection in natural phase space. This step repeats until it finds all substructures. The  Rockstar is efficiently parallel and can be run on high resolution simulations and can be used to analyse the performance of halos .\\
\begin{figure}
	\begin{center}
		\includegraphics[width=0.32\textwidth]{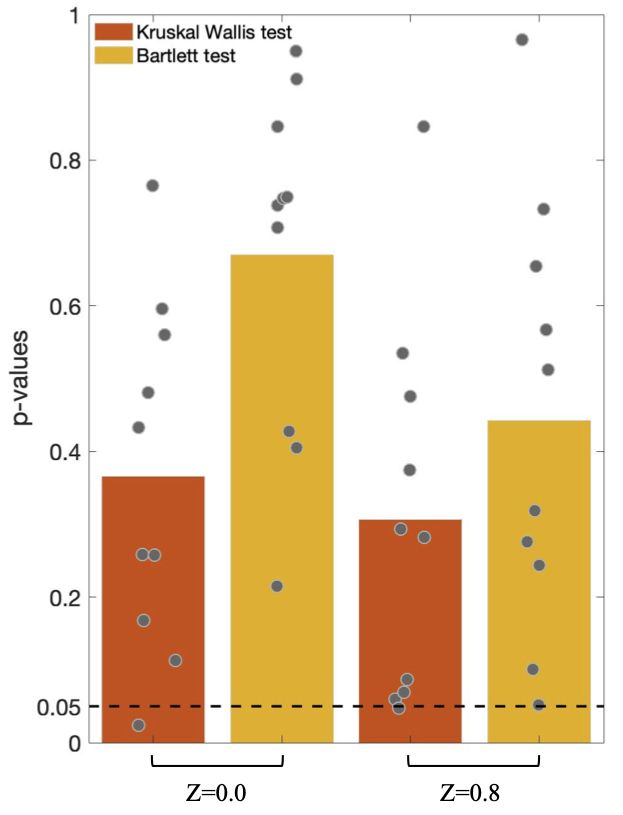}
	\end{center}
	\caption{ The bar-graph shows the homogeneity evaluation  of the velocity dispersion. This exhibits p-values of two statistical tests, Kruskal-Wallis test and Bartlett test, applied to the velocity dispersion of three models obtained via 10 simulations at two redshifts $z=0.0$ and $z=0.8$. }
	\label{Figpopulation}
\end{figure}
\begin{figure*}
\centering
   \includegraphics[width=0.73\linewidth]{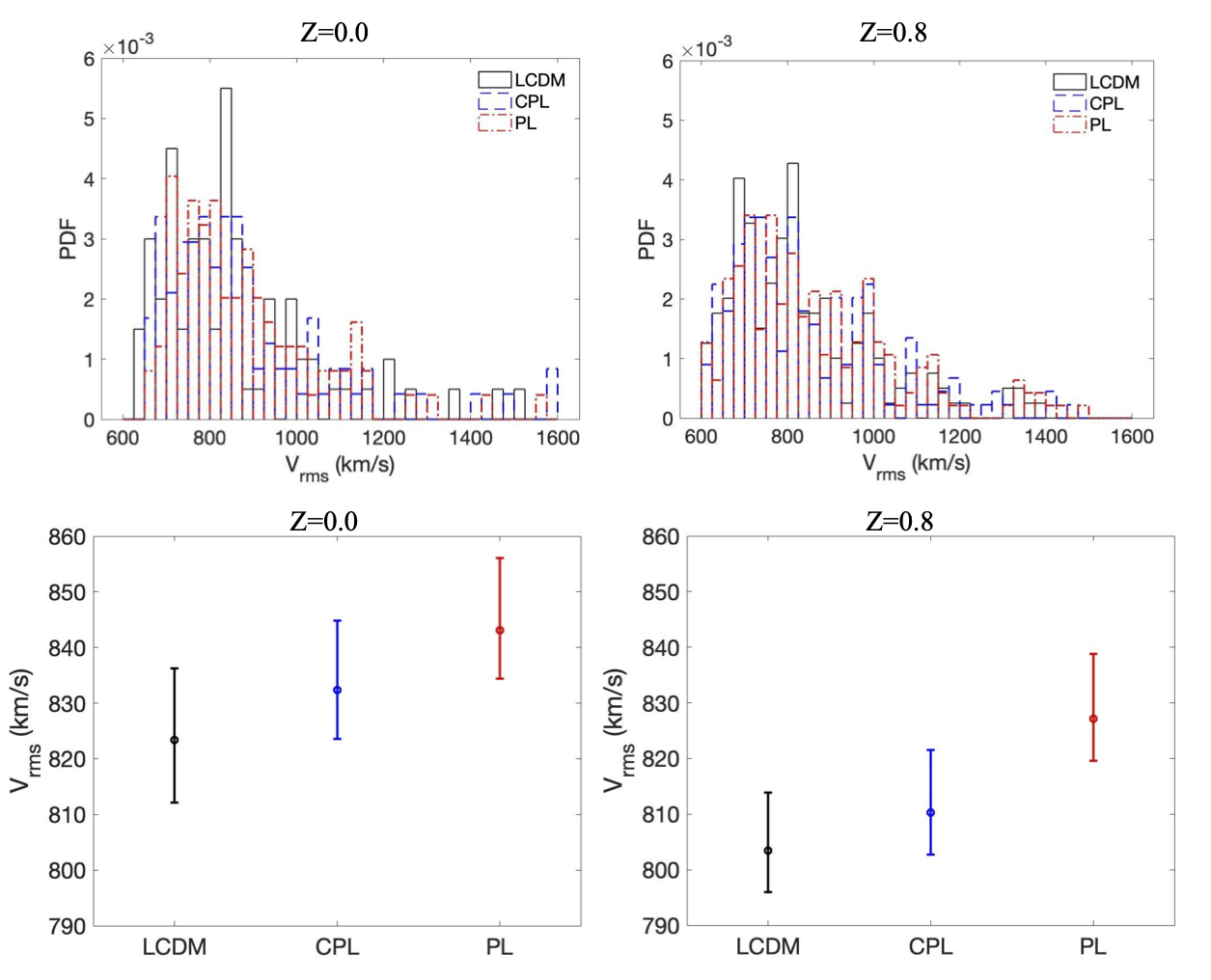}\hfil
\caption{The top row shows the probability distribution functions (PDFs) of the velocity dispersion obtained from the simulation of three models, LCDM, CPL and PL in two redshifts $z=0$ (left panel) $z=0.8$ (right panel). The below plots exhibit the median and interquartile range of the models' density of voids at two thresholds ( $z=0$ and $z=0.8$).
}
\label{V}
\end{figure*}
The Rockstar halo catalogue aims to provide a history for each halo and gives us more information such as position, halos mass, radius, velocity, and velocity dispersion. We found among all properties, velocity dispersion is sensitive to the dark energy models. The fig. \ref{Figpopulation} shows the mean p-value of Kruskal-Wallis test that compared the velocity dispersion, halos velocity, halos mass, and halo radius obtained from three models, $\Lambda$CDM, CPL and PL via 10 simulation boxes at two redshifts with 1 $\sigma$ confidence level. This indicates that among the mentioned properties, the velocity dispersion with smaller p-values is more sensitive in discriminating the dark energy models, despite the fact that the properties of all models were significantly equal. \\
Figure \ref{Figpopulation} compares the p-values of two statistical tests, Kruskal-Wallis and Bartlett, to evaluate the equality of the medians and variances of velocity dispersion between three proposed models at each simulation at two redshifts. No significant difference were observed between calculated p-values (Wilcoxon signed-rank test against $p=0.05$, $p>0.34$). These test confirm that the simulations are homogeneous and we are allowed to select one of the simulations for further analysis of the void finder.\\
Velocity dispersion is defined as the sum in quadrature of the component-wise velocity standard deviations:
\begin{equation}
v_d= \sqrt{\sigma_{vx}^{2}+\sigma_{vy}^{2}+\sigma_{vx}^{2}}
\end{equation}
The probability distribution functions (PDFs) of velocity dispersion for two redshift are shown in the top panels of figure \ref{V}. The below plots exhibit the median and interquartile range of the models' velocity dispersion at two redshifts with 1 $\sigma$ confidence level. Despite the velocity dispersion of three models at two reshshift is not Gaussian, they are equally distributed in the same range of the velocity (Kolmogorov–Smirnov test, $p>0.05$). Moreover, we did not find any statistical difference between the velocity dispersion of two models CPL and PL and the velocities velocity dispersion of the $\Lambda$ CDM when a non-paramteric test, wilcoxon rank sum test, applied to the velocity dispersions to compare their medians (Figure \ref{V}, $p>0.05$). 
The statistical values of the  wilcoxon rank sum test of the velocity dispersion (SWV) of dark matter halos for the given dark energy models with respect to $\Lambda$ CDM. is reported in Table \ref{FFF}.

\subsection{Void Finder Algorithm}
\begin{figure*}
\centering
\includegraphics[width=0.73\linewidth]{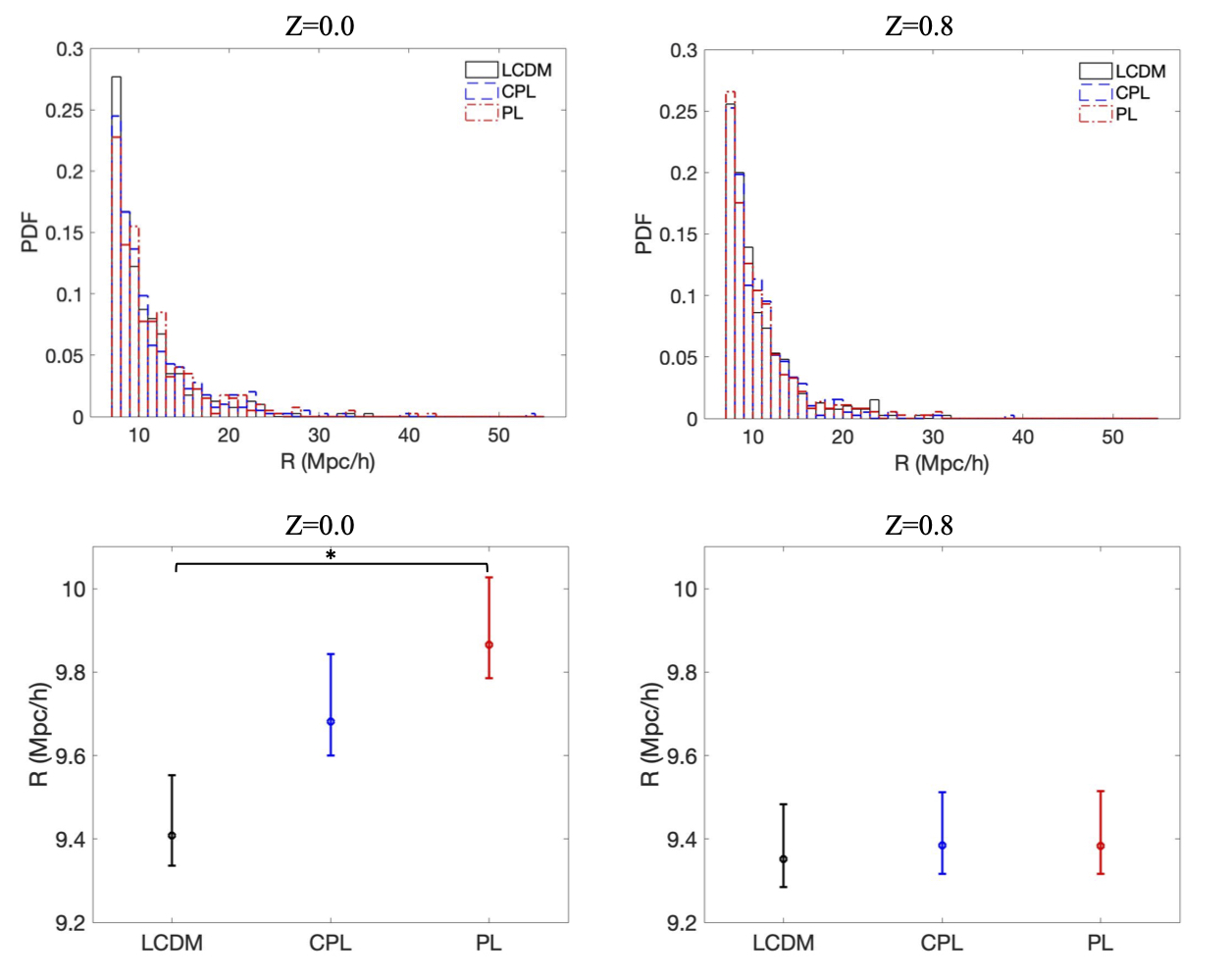}\hfil
 \caption{The top row shows the probability distribution functions (PDFs) of the effective radius of three models, LCDM, CPL and PL at two redshifts $z=0$ (left panel) $z=0.8$ (right panel). The below plots exhibit the median and interquartile range of the models' effective radius at two thresholds (left: $z=0$; right: $z=0.8$), all with 1 $ \sigma $ confidence level. This indicates that only the PL model produced larger void compared with other models ($p=0.02$).}
\label{R}
    \includegraphics[width=0.73\linewidth]{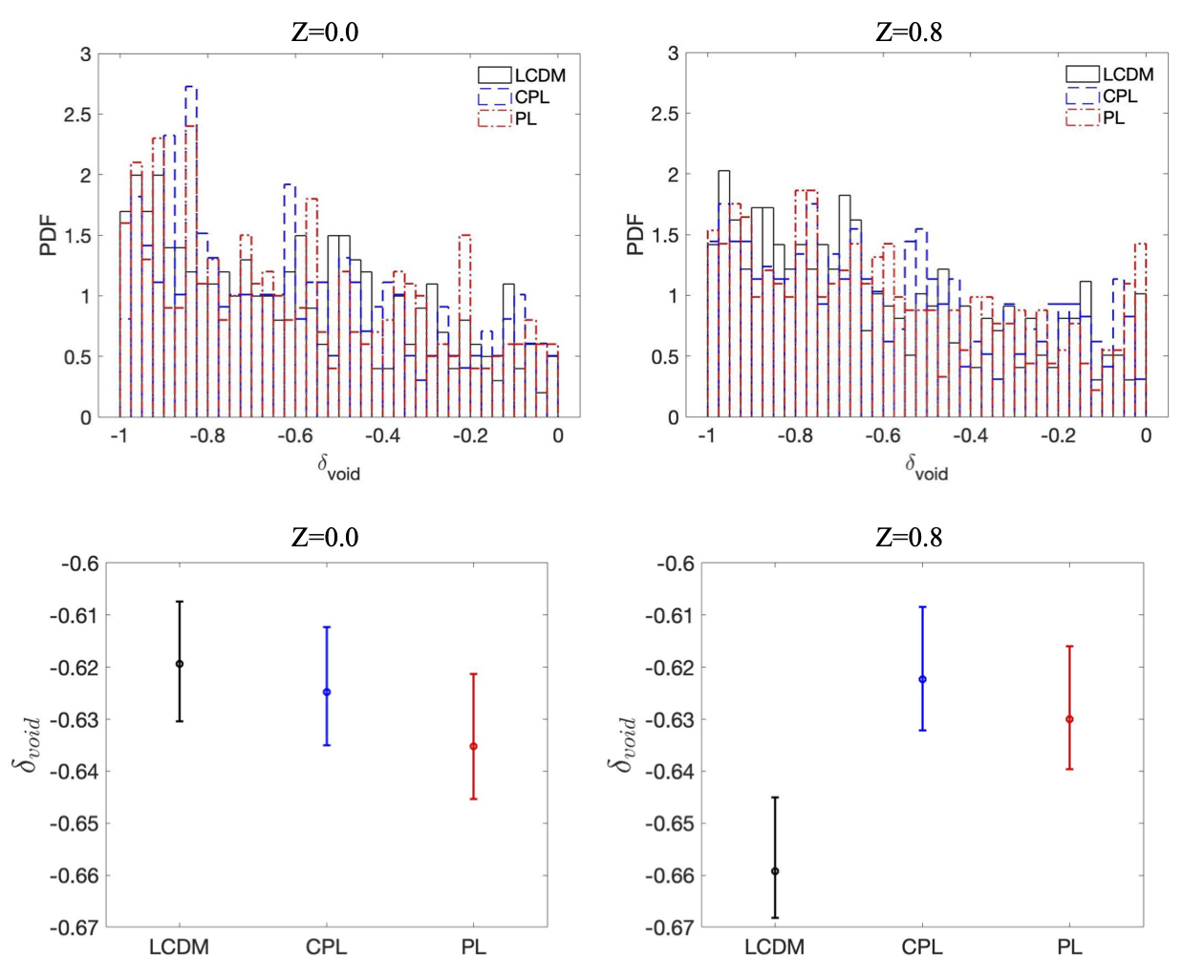}\hfil
 \caption{ The top row shows the probability distribution functions (PDFs) of the density of voids of three models, LCDM, CPL and PL at two redshifts $z=0$ (left panel) $z=0.8$ (right panel). The below plots exhibit the median and interquartile range of the models' density of voids at two thresholds (left: $z=0$; right: $z=0.8$) with 1 $\sigma $ confidence level.}
\label{D}
\end{figure*}

In order to evaluate the potential impact of the models introduced in section 2 on cosmological structures, we compared the properties of voids extracted in the models. 
To do that, we employed the algorithm developed
in  \cite{Tavasoli:2012ai}, which is an updated 3D version of the earlier 2D void finding algorithm Alikio and Mahonen (AM) \cite{Aikio}.
We emphasize that we considered the AM
statistics just as a tools for the relative (not absolute) measurement of some properties of voids (e.g. effective radius and density contrast) in all simulated catalogues. Prior to applying the AM algorithm to the three models, the simulated box sample was girded up to cells of 1 Mpc/h as.
The algorithm works through the first distinguishing between wall and field halos by considering distances to the third nearest neighbour
\cite{hoyle2002voids}. Voids were then identified by the following algorithm. Initially, a Cartesian grid was created and distances from
grid cells to wall halos were identified. Each grid cell was then assigned to a sub void by applying the climbing algorithm \cite{schmidt2001size} to reach the grid cell with the locally-largest distance to
the wall. If the distance between two sub voids was smaller than their distances to walls, they were combined into a single void. Finally, all field halos residing within voids were labeled as void halos.
This generated void catalogue, includes variety of voids in size $R_{v}$, and number density
contrast $\delta_{v}$. The number density contrast of a void is defined by $\delta_{v} = (\rho_{v}-\rho_{m}) / \rho_{m}$ 
where $\rho_{v}$ is given by the ratio of the total number of halos inside a given void by
the volume of that void and $ \rho_{m} $ is mean number density of the samples.
For each void, we defined its effective radius $R_{v}$ as the radius of a sphere whose its volume was equal to the void. In order to avoid counting spurious voids in our catalogue,
 the size of voids was considered to be be larger than $R_{v} > 7 Mpc/h $.\\
The probability distribution functions (PDFs) of effective radius and density contrast are shown in top panel of figures  \ref{R} and \ref{D}  for two redshifts bin while the below plots exhibit the median and interquartile range of the models effective radius and density contrast with 1 $\sigma$ confidence level. Both the effective radius and the density contrast of three models were distributed equally in the same range (Kolmogorov–Smirnov test, $p>0.05$). Although the shape of the distributions are statistically same, the effective radius of the PL models are larger than the radius in $\Lambda$ CDM (Figure \ref{R}, Wilcoxon rank sum test, $p=0.02$). However, no difference was observed in other properties (Table \ref{FFF}). Besides, the statistical values of the Wilcoxon rank sum test
of effective radius (SWR) and density contrast (SWD) of voids of dynamical dark energy model against $\Lambda$ CDM is reported in the table \ref{FFF}.

\begin{table}
	\begin{center}
	\centering
	\caption{ The table shows the statistical value of the  wilcoxon rank sum test of effective radius (SWR), density  of voids (SWD) and the velocity dispersion (SWV) of dark matter halos for the given dark energy models, CPL and PL against  $\Lambda CDM$.}
	\begin{tabular}{|l|c|l|l|l|c|l|}
		\hline
		\hline
		\multirow{3}{*}{Model} & \multicolumn{4}{c|}{Voids}                                                                                        & \multicolumn{2}{c|}{Halos}                             \\ \cline{2-7} 
		& \multicolumn{2}{c|}{SWR}                                & \multicolumn{2}{c|}{SWD}                                & \multicolumn{2}{c|}{SWV}                               \\ \cline{2-7} 
		& z=0.0                      & \multicolumn{1}{c|}{z=0.8} & \multicolumn{1}{c|}{z=0.0} & \multicolumn{1}{c|}{z=0.8} & z=0.0                     & \multicolumn{1}{c|}{z=0.8} \\ \hline
		CPL                    & \multicolumn{1}{l|}{0.34} & 0.92                      & 0.57                       & 0.22                       & \multicolumn{1}{l|}{0.27} & 0.63                       \\ \hline
		PL                     & \multicolumn{1}{l|}{0.022} & 0.79                      & 0.69                       & 0.31                       & \multicolumn{1}{l|}{0.11} & 0.13                       \\ \hline  \hline
	\end{tabular}
    \label{FFF}
    \end{center}
\end{table}

\section{DISCUSSION}\label{discussion}
In this research, we analysed the statistical properties of the dark matter halos and voids as discriminators of two dynamical dark energy models, CPL and PL, against $\Lambda CDM$ model.To analyse the statistical features of the halos and voids, we performed a N-body simulation based on General Relativity \cite{Adamek:2016zes} considering three models,$\Lambda CDM$, CPL and PL \cite{Rahvar:2006tm,Chevallier:2000qy,Ebrahimi:2018eod} for the background evolution of the cosmology. The Rockstar Halo finding algorithm was used to extract the halos of the dark matter for these models. Then, properties of the halos, mass, radius, velocity, and velocity dispersion , were compared between three models. Although no difference between the properties of the halos obtained from the models was not observed (Fig \ref{copmarsim}), the p-values of the velocity depression was smaller compared to other properties. This suggested to consider the velocity depression as a sensitive indicator for further analysis in this study. However, two statistical methods were utilised to evaluate the homogeneity between all simulations (Fig \ref{Figpopulation}). We found that there is no difference between all simulation. This allowed us to select one of these simulation for the next step analysing voids. Thought a searching algorithm developed for this study to find voids in the simulated data (see the Section \ref{result}), different set of voids were obtained for the models. To investigate the properties of the voids in the dark matter, the radius and density contrast of the voids are extracted. we found that although the both radius and density contrast of the models are equally distributed in the form (top Figs \ref{R} and \ref{D}), the PL model garnered larger voids on average (bottom Fig \ref{R}). Since this study reports and compares the properties of the voids and halos generated from the General Relativity base simulation \cite{Adamek:2016zes} for the first time, we suggest properties of voids can provide the information same as the properties of the halos that is historically more common in the cosmology. Moreover, the developed void finder in this study along with the simulation can be considered as a good platform to explore and understand the dark energy properties in future.

Our simulation based on the General Relativity and the void finder algorithm are novel, there are still some limitations in the method of extracting voids' properties. Our simulation was limited at only two redshift bin, $z=0.8$ and $z=0.0$. To track the evolution of halos and voids in the present of dynamical dark energy model, this will be useful to consider more redshifts bin. To speed up the simulation process, we considered 1 MPc/h for the resolution of both simulation and the void finder while this parameter can be varied in higher resolution, but with the expensive computational cost. Indeed, the higher resolutions using parallel processing will result in better discrimination of the voids' radius (Fig \ref{R}). Moreover, the Gas was not included in the N-body simulation we used in this study. This causes a limitation to compare our result with the real observations data. In addition , we modified Genolution on background equations and  did not employ perturbation of dark energy (i.e. in clustering dark energy). Adding the clustering of dark energy will be benefits to get more appropriate results.\\
 Further study is required to complete our simulation with higher resolution N-body code and void finder and adding the clustering of dark energy models to investigate differences of the models respect to $\Lambda CDM$ which is one of the aim of Euclid project \cite{Laureijs2011gra}.

\section{CONCLUSIONS}\label{conclusion}
We examined the dynamical dark energy effect on the voids and dark matter halo properties through N-body simulation in the General Relativity framework by the Gevolution. Our results demonstrate that the effective radius of voids is a better discriminator between the models investigated compared to the density contrast of voids at $ z=0.0 $. Our simulation shows PL model produces larger voids that is promising for further research. Indeed, adding the clustering of dark energy would be a good opportunity to extract differences for future data. We also found that the velocity dispersion of dark matter is the best halo properties among other properties, although, velocity dispersion could not be used as a discriminator of DE models from $\Lambda CDM$ at $ z=0.8 $ and $ z=0.0 $ (see Fig. \ref{V}) with a statistically significant confidence level.\\
We leave investigation of void properties by adding dark energy clustering through N-body simulation and using neural network to further studies.

\section*{Acknowledgements}
The numerical simulations were carried out on Baobab at the computing cluster of University of Geneva. The authors would like to thank HaDi MaBouDi, Marzieh Farhang and Farbod Hassani for their helpful comments during the project.



\bibliographystyle{mnras}
\bibliography{ref} 





\bsp	
\label{lastpage}
\end{document}